\documentclass[twocolumn,preprintnumbers,amsmath,amssymb]{revtex4}
\usepackage{graphicx}% Include figure files
\usepackage{color}
\usepackage{dcolumn}% Align table columns on decimal point
\usepackage{bm}% bold math
\begin{document}
\title{\bf Induced polarization and electronic properties of carbon doped boron-nitride nanoribbons}
\author{J. Beheshtian$^1$, A. Sadeghi$^{2}$, M. Neek-Amal$^{1,3}$\footnote{Corresponding author email: Mehdi.Neekamal@ua.ac.be}
%\footnote{Corresponding author: Mehdi.Neek-Amal@ua.ac.be}
, K. H. Michel$^3$ and F. M. Peeters$^3$ }
\affiliation{$^1$Department of Physics, Shahid Rajaee University,
Lavizan, Tehran 16785-136, Iran.\\$^2$Department of Physics, Basel
University, Klingelbergestrasse 82, CH-4056 Basel, Switzerland.
\\$^3$Departement Fysica, Universiteit Antwerpen, Groenenborgerlaan
171, B-2020 Antwerpen,
 Belgium.}
\date{\today}

\begin{abstract}
The electronic properties of boron-nitride nanoribbons (BNNRs) doped
with  a line of carbon atoms are investigated by using density
functional calculations. Three different configurations are
possible: the carbon atoms may replace a line of boron or nitrogen
atoms or a line of alternating B and N atoms which results in very
different electronic properties. We found that: i) the NCB
arrangement is strongly polarized with a large dipole moment having
an unexpected direction, ii) the BCB and NCN arrangement are
non-polar with zero dipole moment, iii) the doping by a carbon line
reduces the band gap independent of the local arrangement of boron
and nitrogen around the carbon line, iv) an electric field parallel
to the carbon line polarizes the BN sheet and is found to be
sensitive to the presence of carbon dopants, and v) the energy gap
between the highest occupied molecular orbital and the lowest
unoccupied molecular orbital decreases linearly with increasing
applied electric field directed parallel  to the carbon line. We
show that the polarization and energy gap of
 carbon doped BNNRs can be tuned  by an electric field applied
parallel along the carbon line.
\end{abstract} \maketitle

%\section{Introduction}

Single layer hexagonal boron-nitride (h-BN) nanosheets have been
recently  produced and placed on top of  a SiO$_2$ substrate~\cite{10}.
 Several experimental groups have also fabricated free-standing
 h-BN single layers using  sputtering of
controlled energetic electron beams~\cite{11,12}. Unlike graphene, a
h-BN sheet is a wide gap insulator, as is hexagonal BN, which is
   a promising material for opto-electronic technologies
~\cite{14,15}, tunnel devices and field-effect
transistors~\cite{naonoletter2012,APL2011}. Small flakes of h-BN,
i.e. BN nanoribbons (BNNRs), are semiconductors
   when hydrogen-passivated on one of the edges~\cite{20,21,22}. The band gap of BNNR
   can be tuned by using different types of atoms for edge passivation ~\cite{23,24,25,26}.
   The zigzag BNNR has half-metallic properties when the boron edge
   is hydrogen-passivated, and the nitride edge is bare~\cite{23}.
   Under a hydrogen-rich environment, the zigzag BNNR with two-hydrogen-terminated
   edges becomes a ferromagnetic metal~\cite{26}. Moreover, the electronic and magnetic
   properties of the BNNRs could be modulated by applying an extra transversal electric
   field. Zhang \emph{et al} used the local density approximation (LDA)  to study the energy
   gap and showed that it can be  tuned by applying a transverse electric field on a semi-infinite hydrogen
   passivated BNNRs with zigzag or armchair edges~\cite{21}. Similarly by using the
    LDA, Park \emph{et al} found that a transverse electric field decreases the band gap of armchair BNNRs
monotonically while the band gap of zigzag BNNRs either increases or
decreases depending on the direction and the strength of the applied
field~\cite{22}. Additionally, random doping BN with  C atoms leads
to spontaneous  magnetization ~\cite{27,28}. Sai \emph{et al} found
piezoelectricity in a heteropolar nanotube depending on the
chirality and diameter of the nanotube which originates from the
piezoelectric response of an isolated planar BN sheet~\cite{Sai}.

 %Hybrid C-BN nanostructures
%exhibit interesting
%     electronic and magnetic properties, with  half-metallic characteristics
%     even in the absence of an extra electric field ~\cite{29,30,31,32}.

In this study, we report first-principles calculations
 for the BNNRs  which are doped by a
\emph{zigzag line of carbon atoms} in the middle and
 passivated by hydrogen atoms at the edges. We study different
possible positions of the line of carbon atoms with respect to the
BN-lattice and found that piezoelectricity and consequently the
permanent polarization in the BNNRs can be strongly enhanced, tuned
or even eliminated depending on the type of atoms (B or N) that
surrounds the dopants. We found that doping BNNRs by carbon atoms
decreases the band gap while the electric polarization of the doped
BNNRs depends on the type of atoms (B or N) that surrounds the
dopants. The NCB system is polarized in an unexpected way while BCB
and NCN
have zero dipole moments. %We found that the present approach
%based on  hybrid functionals and the LDA approximation.
We find that an external planar electric field applied along the
C-line either reduces or increases the band gap and the dipole
moment depending on its direction.

%This paper is organized as follows. Section II contains the
%essentials of the used model for BNNRs and the method of calculation.
% In Sec. III we present the main results of our paper including: lattice mismatch, electronic charge distribution, electrostatic potential
% polarizability and metallic properties of different studied BNNRs.
% We will conclude the paper in Sec. IV.

\textbf{\emph{Model and Method }}

\begin{figure*}
\begin{center}
\includegraphics[width=0.45\linewidth]{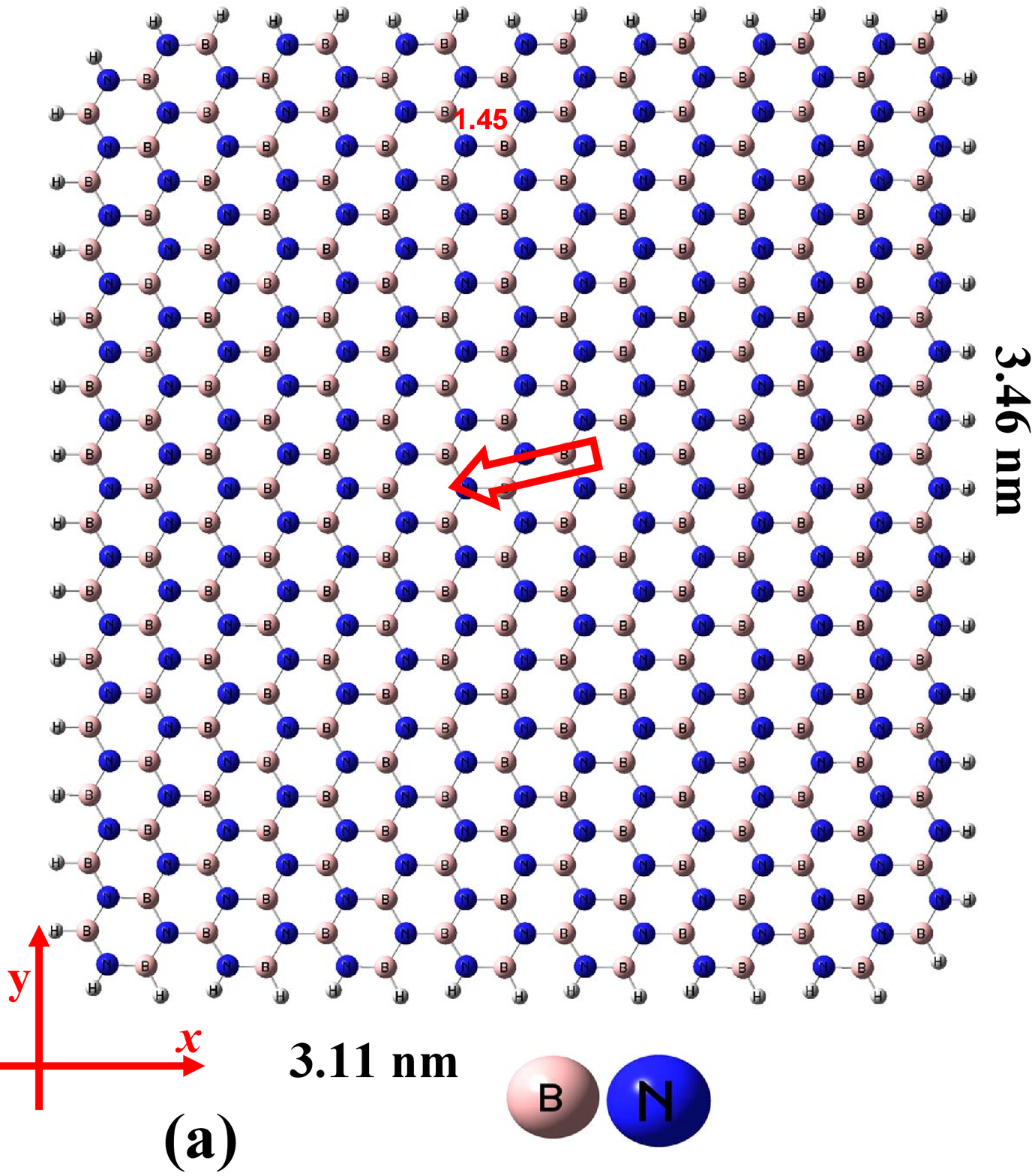}
\includegraphics[width=0.435\linewidth]{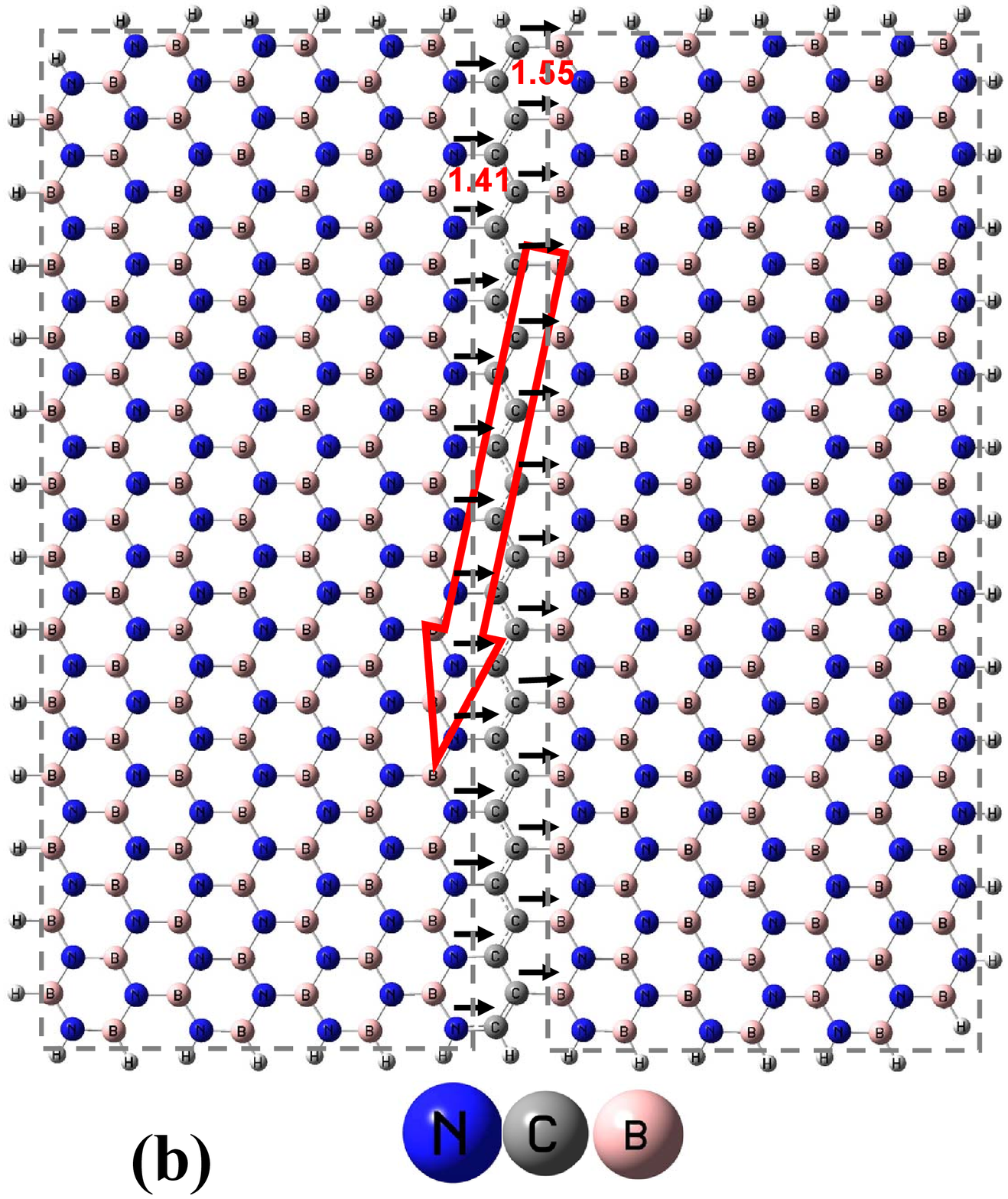}
\includegraphics[width=0.45\linewidth]{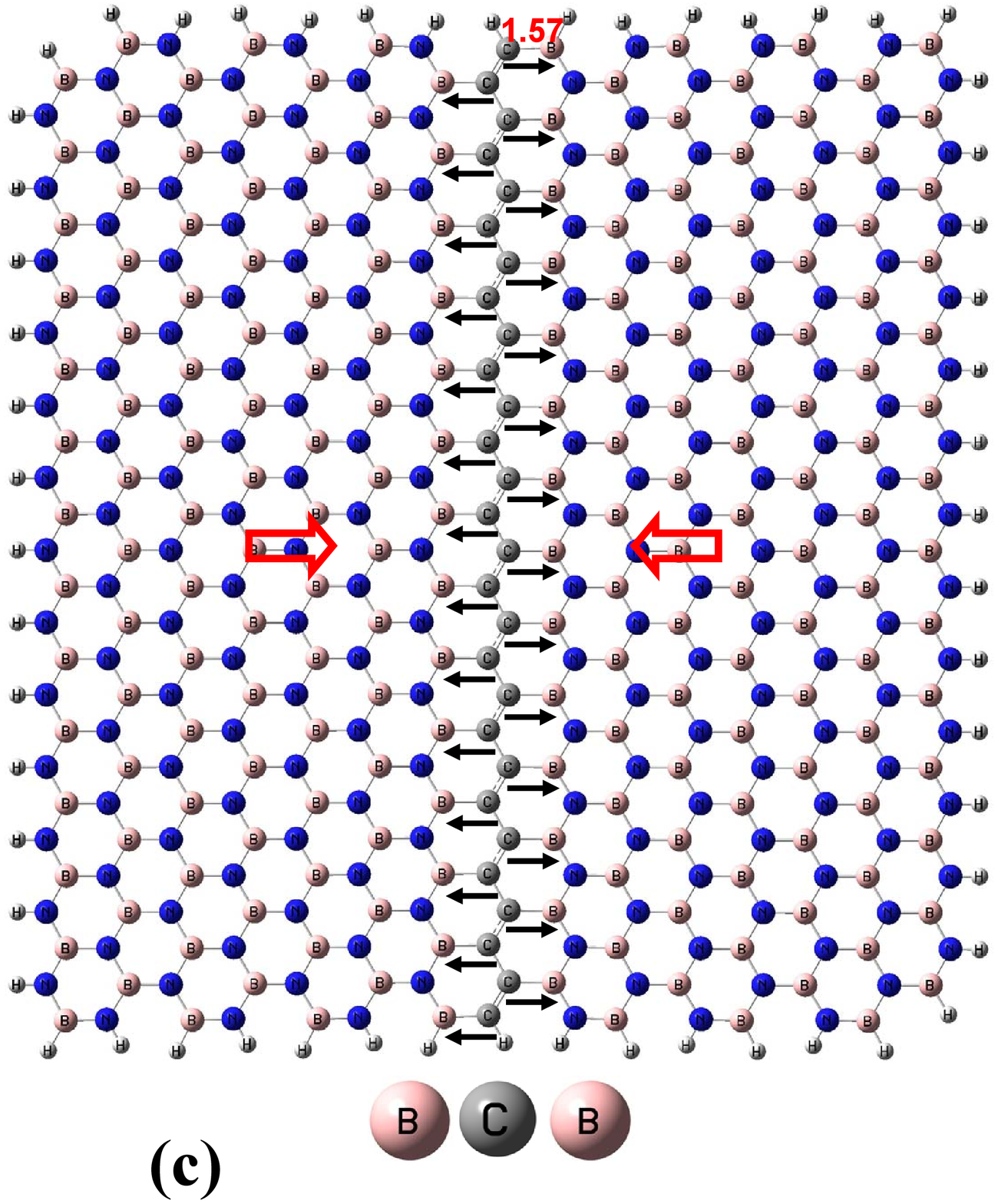}
\includegraphics[width=0.425\linewidth]{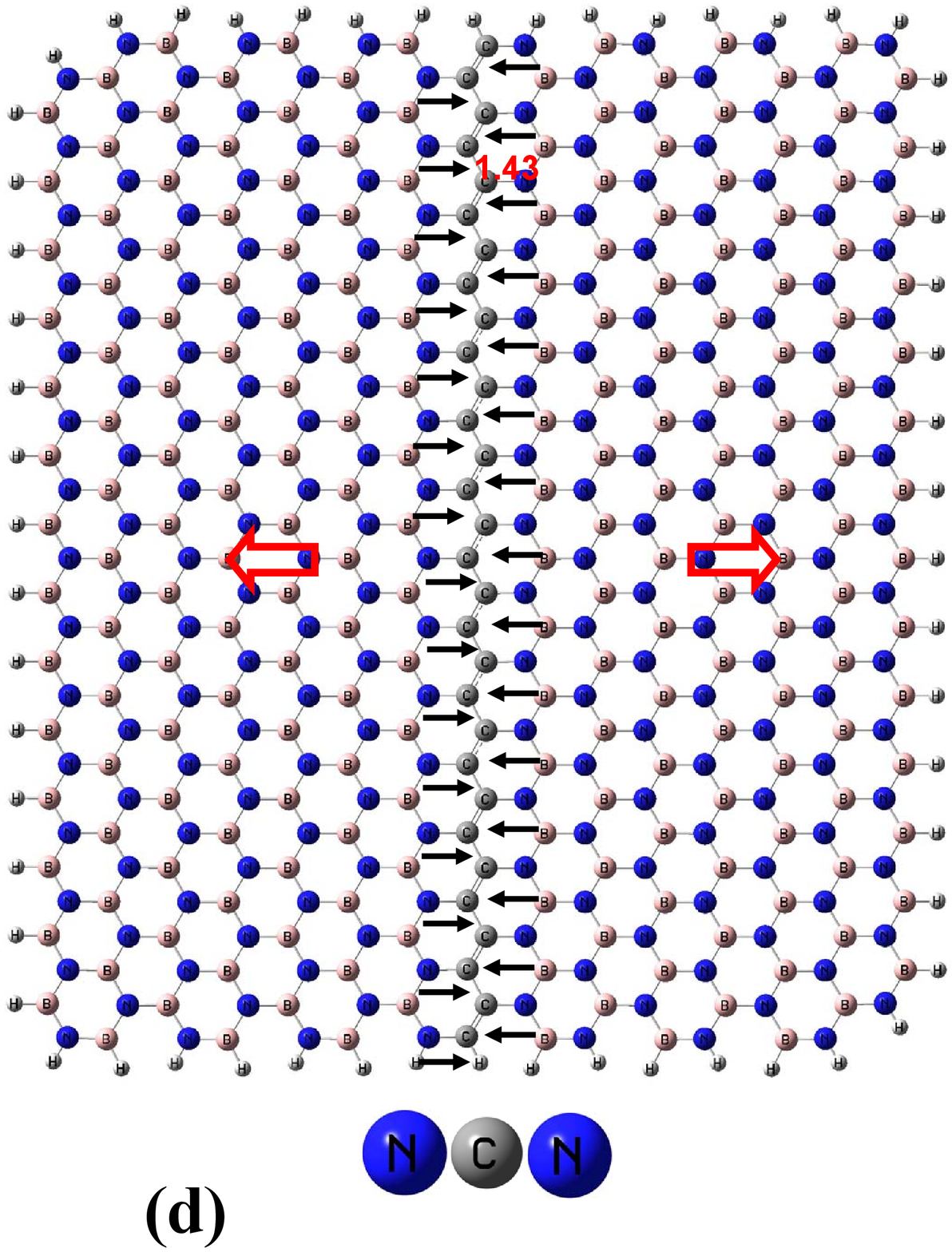}
\caption{(Color online) (a) Perfect boron-nitride nanoribbon (BN).
(b) Boron-nitride nanoribbon with a {zigzag C-line in the middle
surrounded  by  both boron and nitrogen atoms on both sides (NCB).
(c) Boron-nitride nanoribbons with the C-line surrounded only by
boron atoms on both sides (BCB). (d) Boron-nitride nanoribbons with
the C-line surrounded only by nitrogen atoms on both sides (NCN).}
In (a) and (b) red arrows refer to the total dipole moment of the
systems and in (c), (d) red arrows refer to the total dipole moment
on the two sides. The black horizontal arrows refer to the local
dipoles corresponding to the B-C or N-C bonds. The simulated systems
has size 3.46$\times$ 3.11 $nm^2$ (Fig.~\ref{fig0}). Numbers refer
to the bond lengths in \AA~unit. \label{fig0}}
\end{center}
\end{figure*}

 Our system is a finite-size h-BN sheet
which is passivated by hydrogens at its edges. A rectangular h-BN
ribbon $3.1 \times 3.4\,nm^2$ in size is taken as a perfect
nanoribbon, see Fig.~\ref{fig0}.
 Total number of B and N atoms in the sheet is 418, in addition there are 56 H atoms at the
edges. A zigzag line of 28 atoms  is inserted in the middle of the
system replacing a line of h-BN atoms. We have studied three
different positions of the carbon line (C-line) with respect to the
BN-lattice as illustrated in Figs.~\ref{fig0}(b-d) and compare the
results with the undoped sheet (Fig.~\ref{fig0}(a)). Three different
ways of inserting the central C-line are: i) a C-line in the center
with boron and nitrogen atoms on both sides (Fig.~\ref{fig0}(b))
which we name {NCB}, ii) a C-line in the center with boron atoms on
both sides (Fig.~\ref{fig0}(c)) which we name {BCB}, iii) a C-line
in the center with nitrogen atoms on both sides (Fig.~\ref{fig0}(d))
which we name {NCN}. Notice that all these systems are passivated by
hydrogen atoms at the edges to saturate the edge chemical bonds. In
order to have equal sizes on both sides of the C-line, one has to
simulate a non-exact rectangular flake, i.e. the right-up and
down-left corners are not orthogonal, as illustrated by the two
dashed rectangles in Fig.~\ref{fig0}(b) having  the same size.
 Therefore in the RHS (LHS) rectangles in NCB, BCB and
NCN there are additional  B(N), B(B), N(N) atoms, respectively (see
Figs.~\ref{fig0}(c-d)).

We employ density functional theory as implemented in GAUSSIAN (G09)
~\cite{34} which is an electronic structure packages that uses a
basis set of Gaussian type of orbitals. For the exchange and
correlation (XC) functional, the hybrid functional B3LYP is adopted
in G09. For some particular cases, and for comparison purposes, we
also used the LDA  as implemented in the BigDFT ~\cite{bigDFT}
package which uses real-space wavelets. In both cases, the self
consistency loop iterates until the change in the total energy is
less than $10^{-7}$ eV, and the geometries are considered relaxed
once the force on each nucleus is less than 50\,meV/\AA.~Using the
6-31G* basis set in G09, we expect that our calculation is capable
to provide a reliable description of the electronic properties of
the different systems.  Natural bonding orbital (NBO)
analysis~\cite{33} is also performed at the B3LYP/6-31G* level of
theory~\cite{34}.

%To study the polarization effects in the presence of an
%external electric field and in order to have an
%independent confirmation of our results we used also the BigDFT.
%The latter is
%efficient and massively parallel electronic structure code.
%Calculations of both the Kohn-Sham electronic orbitals and the
%electrostatic Hartree potential for an isolated system is done in
%BigDFT without the need to choose a supercell~\cite{Poisson}. Thus no artifact
%arises when an electric field is applied. The local density
%approximation (LDA) is used as the XC functional while the
%nuclei-electrons interaction is approximated by HGH
%pseudopotentials~\cite{HGH}. The grid resolution which in turn
%determines the number of basis functions is chosen such that an
%accuracy  of 1~meV per atom is achieved in energy. When the electric
%field is changed, the BNNR is  allowed  to relax again  with the
%same criteria mentioned before.
%

\textbf{\emph{Lattice mismatch }}

\begin{figure*}
\begin{center}
\includegraphics[width=0.75\linewidth]{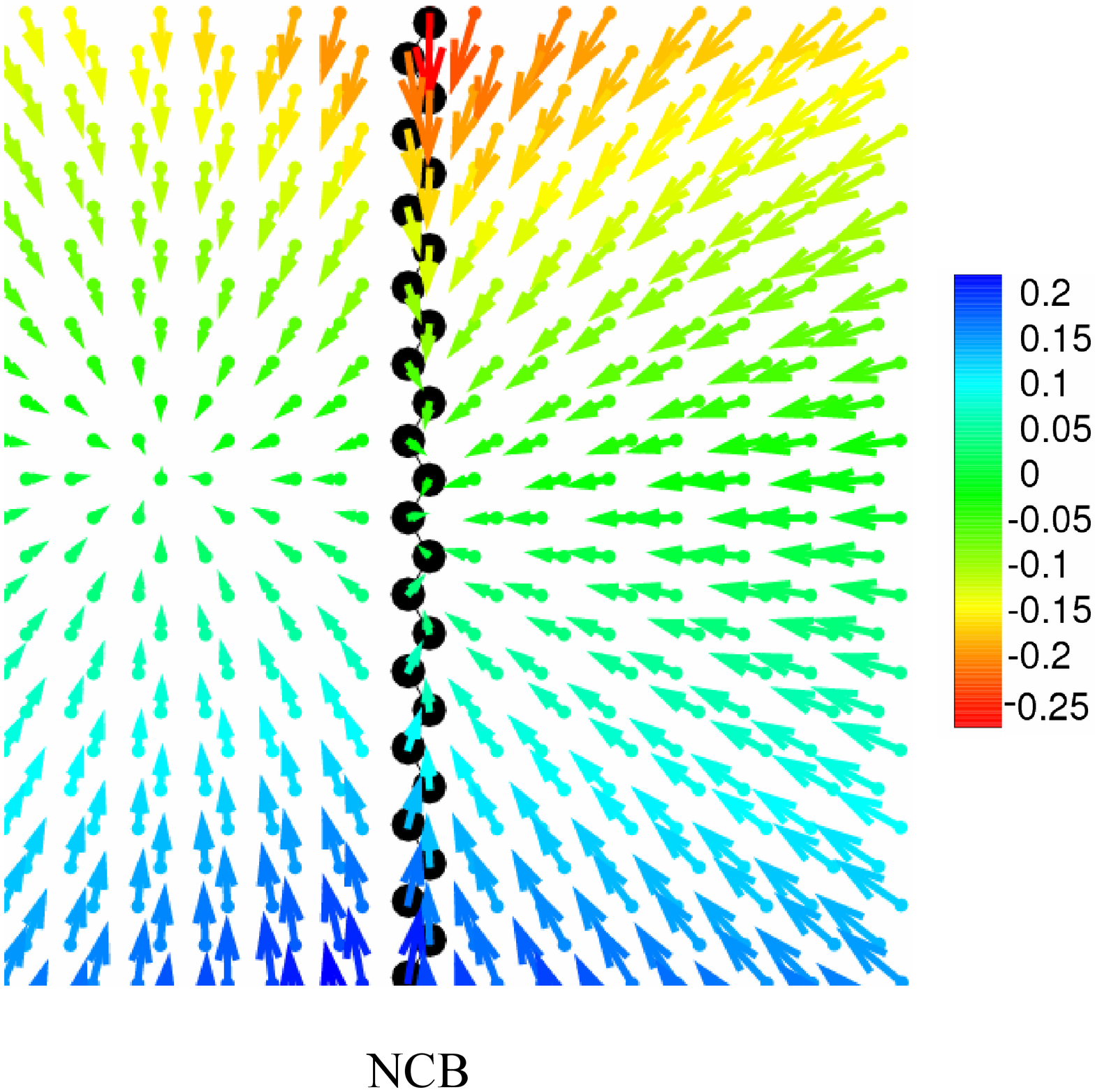}
\caption{(Color online) Atomic displacement after inserting an
atomic carbon line in the middle of the BN sheet, i.e. NCB. The
color coding indicates the y-component of the displacement vector,
i.e. red (blue)$\equiv$ downward (upward) displacement.
\label{figdis} }
\end{center}
\end{figure*}

We recall that in bulk h-BN the in-plane B-N bond length
$d_{BN}=1.446$\AA~has been determined by X-ray
scattering~\cite{pease52}. Early ab-initio
calculations~\cite{kern99} yielded $d_{BN}=$1.35\AA.~We obtain
$d_{BN}=$1.45\AA~which agrees much better with experiment. On the
other hand the bond lengths $d_{NC}$ and $d_{BC}$ between the
inserted C-line atoms and the N and B surrounding atoms, as well as
the intra-chain bond length $d_{CC}$ between C-atoms, depend on the
specific configuration of the atoms in the NCB, BCB and NCN systems
(see Fig.~\ref{fig0}). We summarized the results of our calculations
in Table I. We found for the three doped BNNRs that
$d_{CC}<d_{BN}<d_{BC}$. Notice that the relative sheet bond lengths
$d_{NC}=$1.41\AA~ and 1.43\AA~ in NCB and NCN, respectively, are an
indication of the partially double bond character of the N-C bonds
at the central C-line is remarkable. The difference in bond lengths
around the C-line in NCB are clearly visible in Fig.~\ref{fig0}(b).
The values $d_{BC}=$1.55\AA~ and $d_{NC}=$1.41\AA~ lead  to a
lattice mismatch and symmetry breaking between the left and the
right part of our system along the inserted C-line. In
Fig.~\ref{figdis}, the arrows show the corresponding atomic
displacement pattern for NCB. On the other hand the systems BCB and
NCN (Figs.~\ref{fig0} (c,d)) are symmetric with respect to the
C-line. The changes in the bond-length and the dependence of the
configuration on the interface leads to changes in the electronic
polarization and consequently to a piezoelectricity effect. The
results of this electro-mechanical coupling will be discussed in the
following part of this section.

\textbf{\emph{Electronic charge distribution}}

The perfect infinite BN sheet is a neutral sheet with a uniform
charge distribution far from the edges. Indeed, in an infinite two
dimensional h-BN sheet the point group symmetry is D$_{3h}$ and
there is no permanent dipole moment. However in finite BNNRs
structure the symmetry is broken which allows piezoelectricity, i.e.
the appearance of a dipole under mechanical deformation. Therefore,
the h-BN sheet is the most simple crystalline structure which allows
piezoelectricity~\cite{Karl2009,Karl2011}. On the other hand, in
BNNRs the finite extension of the sheet and the passivation of the
boundaries by H atoms lead to the appearance of a permanent dipole
moment. For the case of BN (see Fig.~\ref{fig0}(a))
 using B3LYP we obtain a dipole moment $\overrightarrow{P_0}=(-10.58,-1.98)$ D, (in this study the electric dipole moment
  vectors point from the negative charge to the positive charge, e.g. see red
arrow in Fig.~\ref{fig0}(a)), i.e. the system becomes piezoelectric.
In the following we investigate
 the influence of carbon doping on the polarization in BNNRs.

Inserting of  C-line alters the charge distribution mostly in the
neighborhood of the C-line (see Table I), while further away the
charge distribution approaches the one of a perfect BNNR. The
different values of electronegativity for $B(\approx 2.0)$,
$C(\approx 2.6)$ and $N(\approx 3.0)$ largely account for the charge
redistribution and the concomitant formation of local dipoles.

For the NCB  the local dipoles for B-C and C-N bonds around the
carbon line are directed to the right~(Fig.~\ref{fig0}(b))
 which polarized NCB. Surprisingly, although local dipoles are in the
 $x$-direction, we found that the total dipole moment is mainly in
the  $y$-direction (parallel to the carbon line), i.e.
 using B3LYP we find $\overrightarrow{P_0}=(-9.37,-42.89)$D. We attribute this effect
 to the presence of carbon atoms in the middle of
Fig.~\ref{fig0}(b). The C atoms which are bonded to the B (N) atoms
absorb (give) electronic charge yielding two different group of C
atoms, i.e. those bonded to B atoms with negative electronic charge
and those bonded to N atoms with positive electronic charge. This is
the main reason for the unusual direction of the dipole moment. In
fact  the magnitude of $P_y$ is related to the carbon line. Notice
that this is seen only when different types of atoms surround the
C-line, i.e.
the BCB and NCN systems have small $P_x$ and $P_y$. %Therefore, one expects that the
%insertion of another C-line (not connected to the present C-line)
%increases the spontaneous $y-$component of the dipole moment, $P_y$.

As seen from  Figs.~\ref{fig0}(c,d) in the BCB and NCN systems the
dipoles of the left and right hand sides of the C-line are in
opposite directions and no overall dipole moment in the
$x$-direction arises from these edges. There is a small dipole
moment $P_y$ because the corners are differently terminated at the
left and right edges.

 A similar description is possible for the strong local dipoles  on the B-C and
N-C bonds induced by the insertion of  the C-line. Although having a
C-line in the middle, the BCB and NCN systems have vanishing dipole
moments in the $x$-direction
 because the local dipoles on the B-C and N-C bonds around the
 C-line are directed oppositely. The direction of the
local dipole moments, indicated by the horizontal black arrows in
Figs.~\ref{fig0}({b-d}), is chosen according to  the
electronegativity differences of the atoms as discussed earlier. It
is important to note that the number and arrangement of hydrogen
atoms are the same in all model systems, thus the main reason of the
strong total polarization is the local arrangement of B and N atoms
 around the C-line and not the presence of the H-atoms of the edges.

\textbf{\emph{Electrostatic potential }}

\begin{figure*}
\begin{center}
\includegraphics[width=0.755\linewidth]{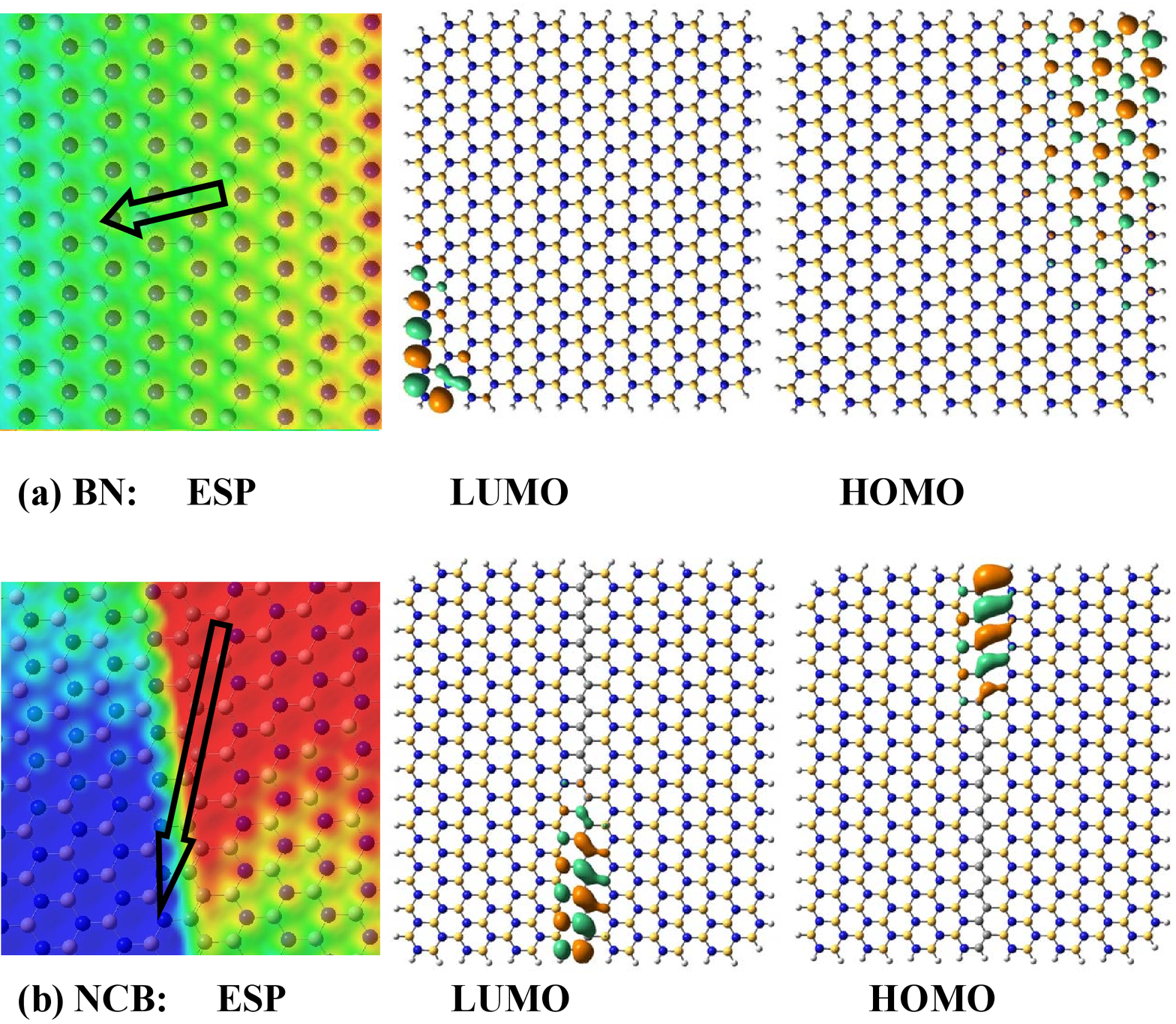}
\includegraphics[width=0.755\linewidth]{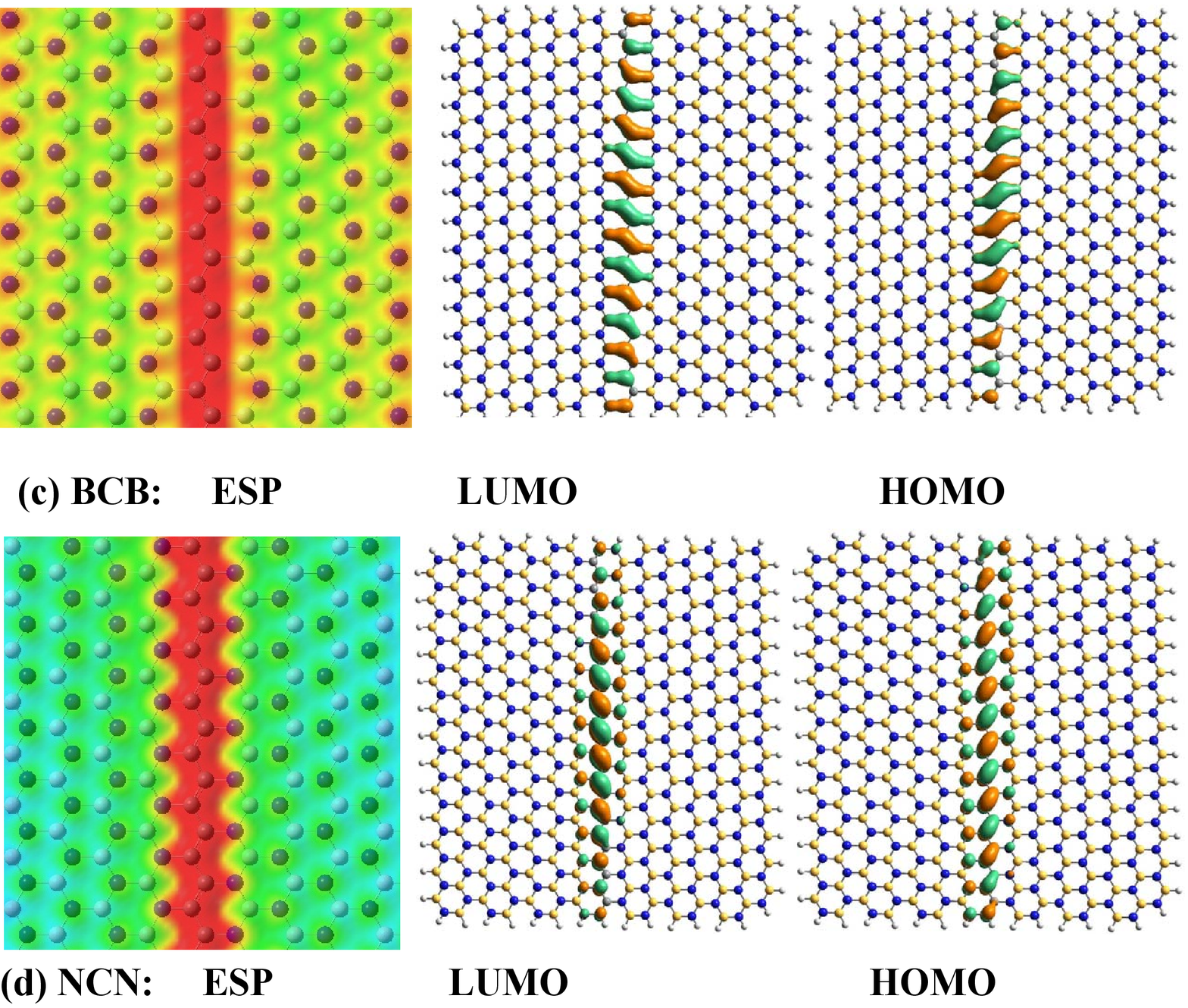}
\caption{(Color online)  Contour plots of the electrostatic
potential (ESP) for the four  systems of Fig.~\ref{fig0} and the
corresponding highest occupied molecular orbitals (HOMO) and the
lowest unoccupied molecular orbital (LUMO). In (a) and (b) the black
arrows refer to the total dipole moment.  \label{figESP} }
\end{center}
\end{figure*}

 The electrostatic potential (ESP) mapped on the plane
of  the h-BN sheet for the representative models is shown in
Fig.~\ref{figESP}.  The sign of ESP at any point depends on whether
the ESP due to  the nuclei and the electrons is dominant at that
point. The difference in electronegativity between the B and N atoms
results in a combination of weak ionic and covalent bonding in the
h-BN sheet. Fig.~\ref{figESP}(a) shows that the B-N bonds have a
larger ESP (blue spots) around the B ions
 as compared to the N sites (red spots).
It shows that the N ions attract electronic charge from the B ions
as one expects in view of the atomic electronegativities.
 A  characteristic pattern of alternating positive and negative ESP regions
is seen over the BN  surface in which the left side has generally
less ESP than the right side. Accordingly, this explains why the BN
 has a permanent dipole moment in accordance of
 the electronegativity differences  on H-B and H-N bonds at the edge  which saturate the dangling bonds.
The NCB, as shown in Fig.~\ref{figESP}(b),  has a steep gradient in
the ESP near  the C-line which results in a strong dipole in this
region which confirms the direction of the dipole moment shown in
Fig.~\ref{fig0}(b). The big arrows in Figs.~\ref{figESP}(a,b) refer
to the dipole moments. Similarly, the direction of the opposite
dipoles in Figs.~\ref{fig0}(c,d) can be well explained by the ESP
variation  in Figs.~\ref{figESP}(c,d). Strong variation of the ESP
around the carbon lines causes strong dipoles but directed
oppositely. In NCB, the antisymmetrical distribution is also due to
the chosen non-perfect geometry of the corners.

\textbf{\emph{Polarizability}}

\begin{figure*}
\begin{center}
\includegraphics[width=0.9\linewidth]{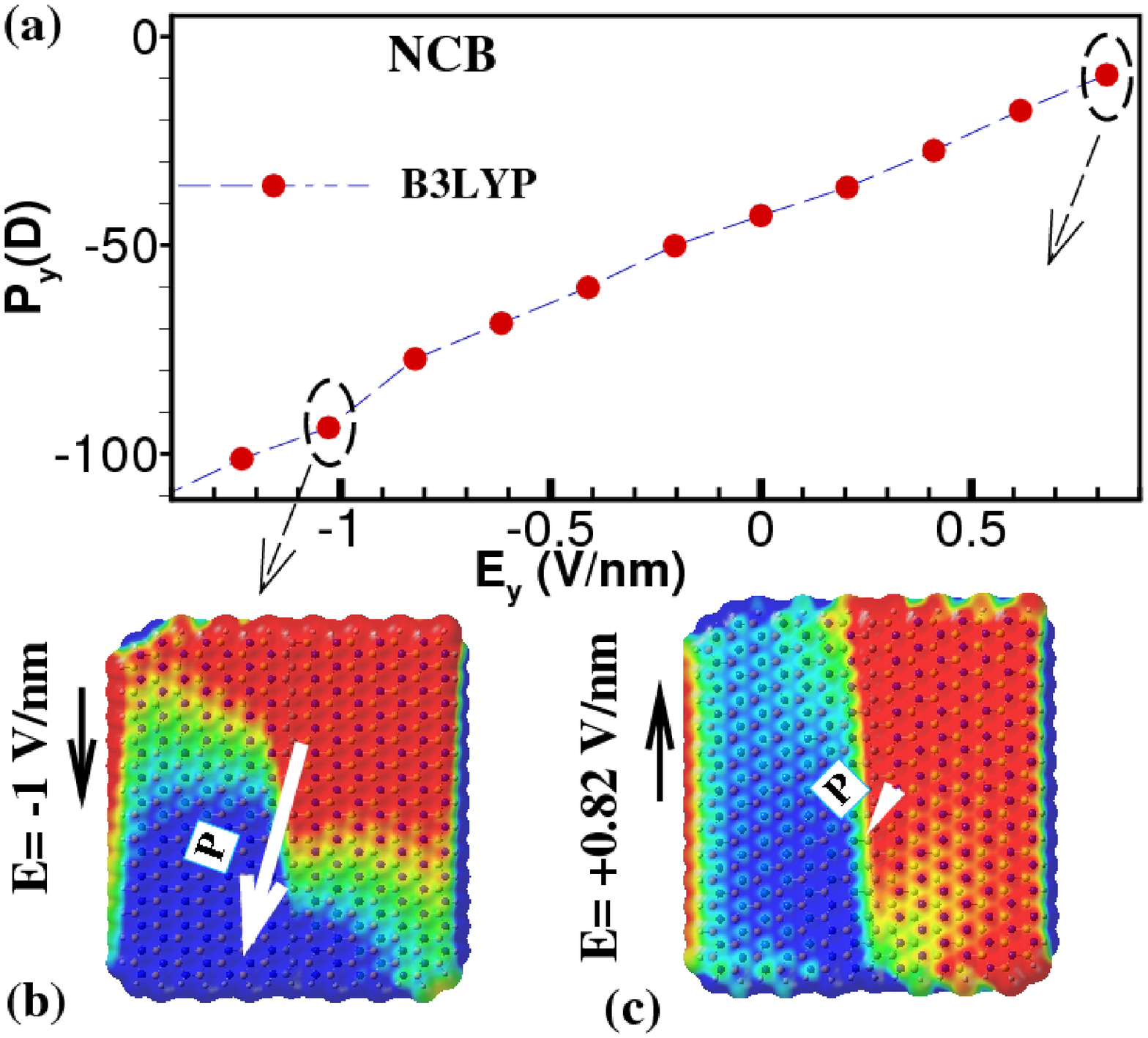}
\includegraphics[width=0.45\linewidth]{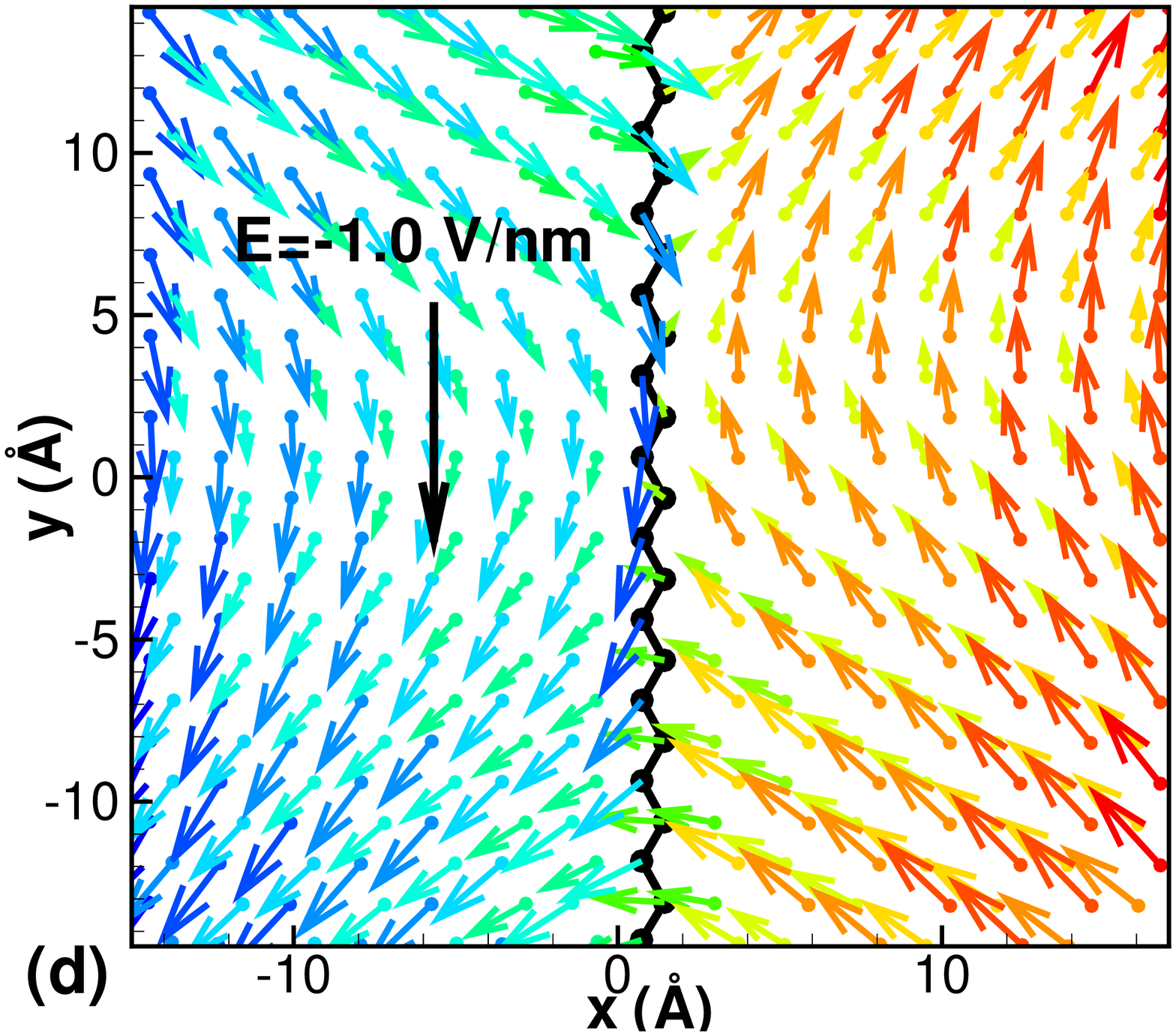}
\includegraphics[width=0.42\linewidth]{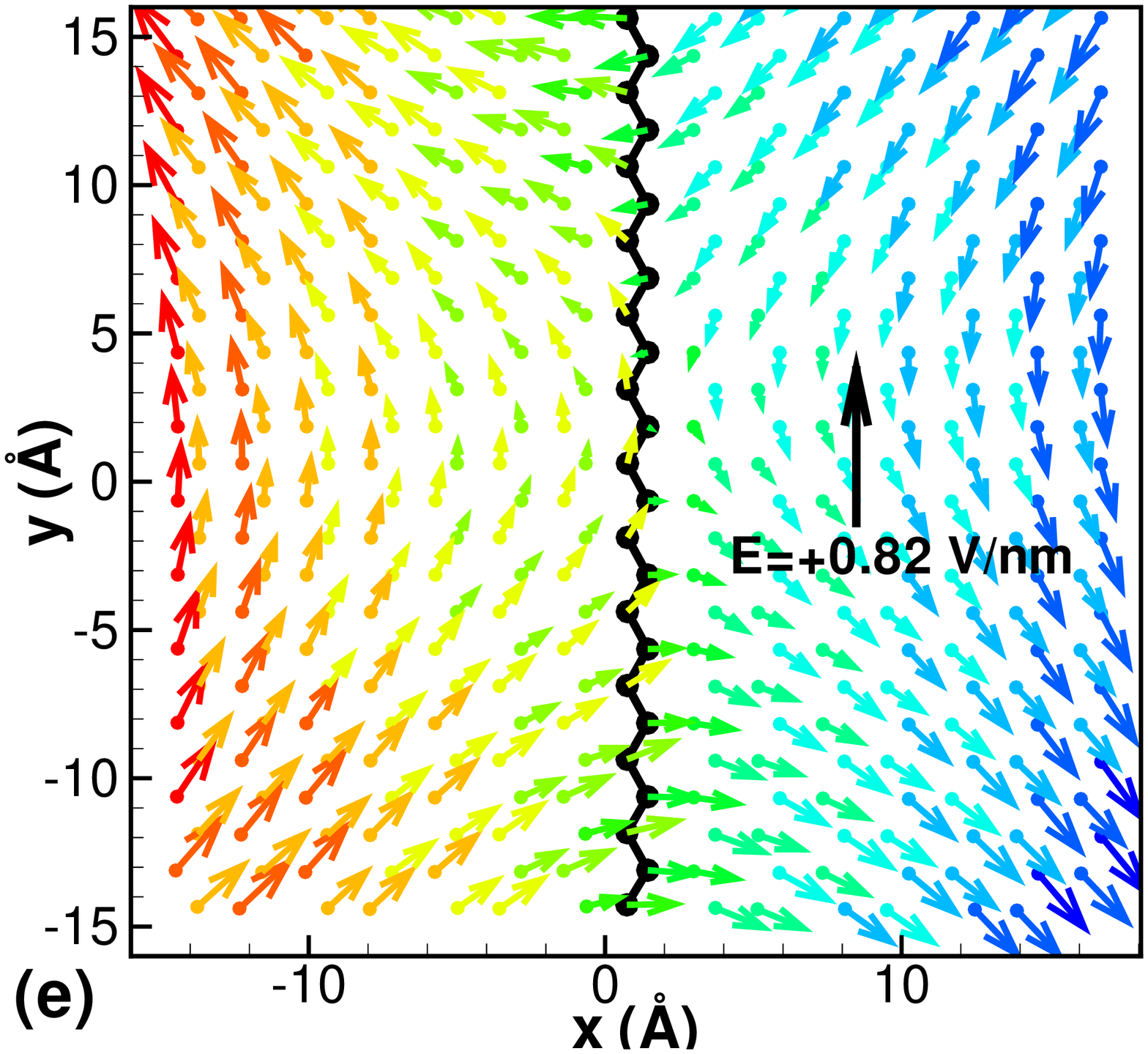}
\caption{(Color online) (a) Electric dipole moment versus applied
electric field, both in the $y$-direction, for the NCB system using
B3LYP functional. The corresponding ESPs and lattice deformation for
$E=E_y$=-1.0 V/nm (b,d) and $E=E_y$=+0.82 V/nm (c,e). The black
arrows (symbols) refer to the direction of the applied electric
field (carbon atoms) and the white arrows refer to the total dipole
moment.  \label{figPE} }
\end{center}
\end{figure*}

 The polarization of the BNNR system varies  in response to an external
electric field. Here we focus on the NCB system which has an
appreciable permanent polarization. We apply a uniform electric
field  in the plane of the flakes parallel to the C-line
($y$-direction)
 and allow them to relax each time the electric field is altered.
The reason for directing the field along $y$ is that we want to
compensate the largest component of the spontaneous dipole moment
$P_0$, i.e. $P_y$. The field strength $E\equiv E_y$ is then
increased step-wise until the induced dipole moment completely
compensates $P_y$. We denote the compensating field strength by
$E_0$. We also inverted $E_y$ to check the linear dependence. Red
symbols in Fig.~\ref{figPE} (for NCB system) show that the dipole
moment has almost a linear dependence on $E$ within the studied
range (-1,0.8)\,V/nm.  The slope of this line gives the
polarizability $\alpha_{yy}=dE/dP_y$. % We carried out
%several additional calculations using BigDFT for compensating $P_y$
%in NCB system. As we see from blue symbols in Fig.~\ref{figPE}(a),
%although LDA approximation gives reasonable dipole moment for
%$E_y=0$ (see Table~\ref{table2})
% as compared to the B3LYP results, but it underestimate (overestimate, i.e. two times larger)
%compensating electric field (dipoles moment).
From B3LYP we found $\alpha_{yy}$=45.66 D nm/V and  $E_0\approx$ 1
V/nm.
% and from LDA
%they are around 85.5 D nm/V and 0.05 V/nm, respectively.
In Figs.~\ref{figPE}(b,c)  we show the  ESP and $P_0$  for
$E_y$=-1.0 V/nm and $E_y$=+0.82 V/nm, respectively. The lattice
deformations due to applied electric field are shown in
Figs.~\ref{figPE}(d,e). It is interesting to note that a positive
 electric field shifts the RHS-atoms to the bottom and
the LHS-atoms to the top, and vise versa for negative field. This
yields  a shear stress along the C-line (filled black symbols). The
larger the electric field the larger deformation and the larger the
enhancement of piezoelectricity.

\emph{\textbf{Metallic Properties }}

\begin{figure*}
\begin{center}
\includegraphics[width=0.9\linewidth]{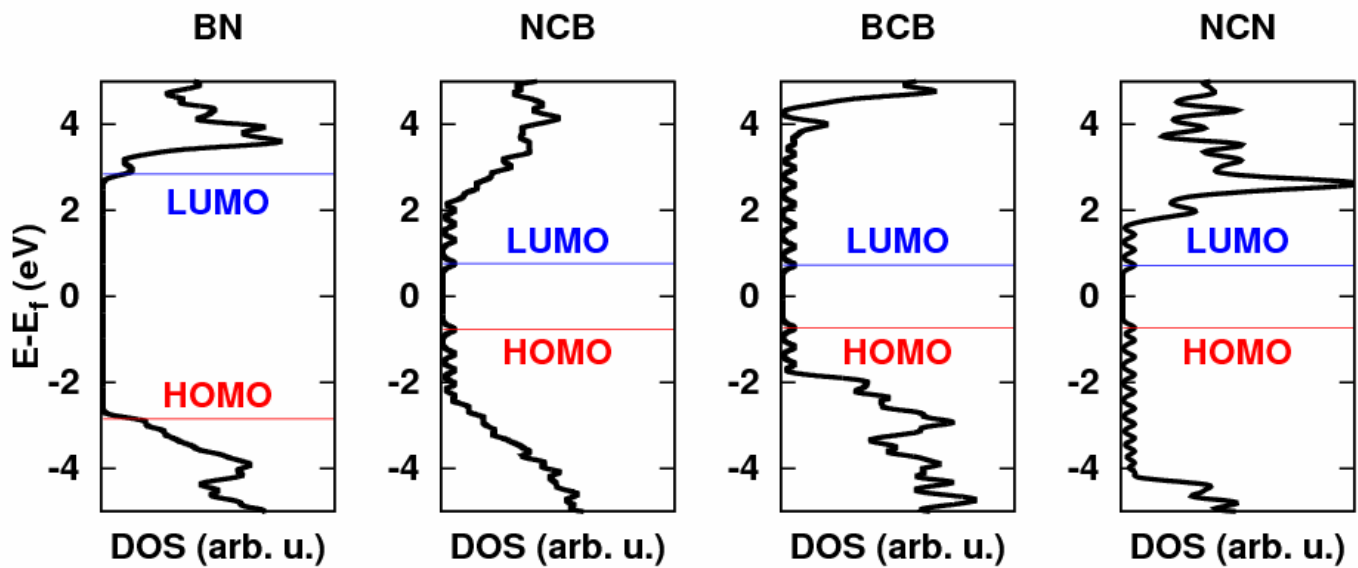}
\caption{(Color online)  Density of states (DOS) for the four model
systems. The presence of the carbon lines in the middle of the
systems decreases the gap and modifies the DOS in different ways for
each particular cases. \label{figdos} }
\end{center}
\end{figure*}

 To show the effect of the insertion of the C-line on the conductivity of BNNRs,
we have listed the calculated energy gaps for our systems in Table
~\ref{table2}. The  gap is calculated as the energy difference
between the highest occupied molecular orbital (HOMO) and the lowest
unoccupied molecular orbital (LUMO).
 It is well known that LDA (which we used in our BigDFT calculations)
  underestimates the energy of the excited levels and therefore the HOMO-LUMO  gap.
  The hybrid functional B3LYP is expected to give a better estimation.
 However,  both approximations reveal the same trend i.e. significant decrease in
the HOMO-LUMO gap after insertion of the C-line. It implies that the
electric conductivity is increased in the doped sheets, however
their gaps are still  large  and therefore the system is not
metallic. We will show later that the HOMO-LUMO gap can also be
tuned by an electric field. Figure~\ref{figdos} illustrates the
density of states (DOS) for the studied BNNRs according to B3LYP.
Note that the insertion of a C-line generates a few new electronic
states within the wide gap between the HUMO and LUMO of the perfect
BN, reducing the gap width significantly.
 Notice  that the distribution of energy levels around the HOMO-LUMO levels in BCB and
NCN are different  and the energy levels for NCN (BCB) are
significantly separated below (above) the HOMO (LUMO).

  The change in the metallic properties can be explained by looking at the distribution
   of  the electronic states near the Fermi level. As seen from Fig.~\ref{figESP} in the BN system the HUMO
  (LUMO) is localized at two corners around the N (B) atoms.
In the BN system, the HOMO is localized at the right-down side with
zig-zag edges while the LUMO is localized at the left-upper corner
with zig-zag edges where these corners are orthogonal opposite to
the other corners, i.e. right-up and down-left with zig-zag edges.
%This particular pattern for HOMO-LUMO yields the left-down directed
%dipole in the BN system.
It shows that both chirality and the local geometry of the edges are
important parameters for localizing the electrons. On the other hand
in the doped systems the HOMO and LUMO states are localized around
the C-line over either B-C or N-C bonds, see
Figs.~\ref{figESP}(b,c). In BCB and NCN systems the C-line opens a
wide channel of few atoms through which the electronic charge can be
transmitted. But the conductivity in the direction perpendicular to
the C-line is expected to be small because no contribution from HOMO
and LUMO is extended along this direction. Notice that the localized
orbitals in BCB and NCN have different patterns. In BCB both HOMO
and LUMO are localized on the C atoms and the adjacent B atoms (i.e.
on the B-C bonds), while the N atoms are not contributing. This is
apparently in contrast to the belief that the HOMO (LUMO) are
localized on N (B). This is due to the larger electronegativity of C
as compared to B. In the NCN system HOMO and LUMO are localized
mainly on the C-C bonds (with different orientation for HOMO and
LUMO) and are partially localized on the adjacent N atoms and the B
atoms never contribute.

   The special shape of the localized HOMO and LUMO states around the two ends of the C-line
   for the NCB is of most interest. The HOMO orbitals in NCB  are localized over a few top C-C bonds and neighbor B
   atoms while for the LUMO we notice that they are localized over a few bottom C-C bonds where B (N) atoms
    in the RHS (LHS partially)  are mainly  contributing with different shape as compared to the HOMO.
It is important to emphasize that: i) we found the same orbital
distribution in a BN system when it is doped by an arm-chair line of
carbon atoms in the middle (those results are not reported
 in this paper), and ii) we did not find such orbitals
employing semi-infinite NCB, i.e. a ribbon with periodic boundary
along the C-line. The special shape of HOMO and LUMO orbitlas found
in NCB are not found in the BN system~\cite{22,21}.

\begin{figure*}
\begin{center}
\includegraphics[width=0.45\linewidth]{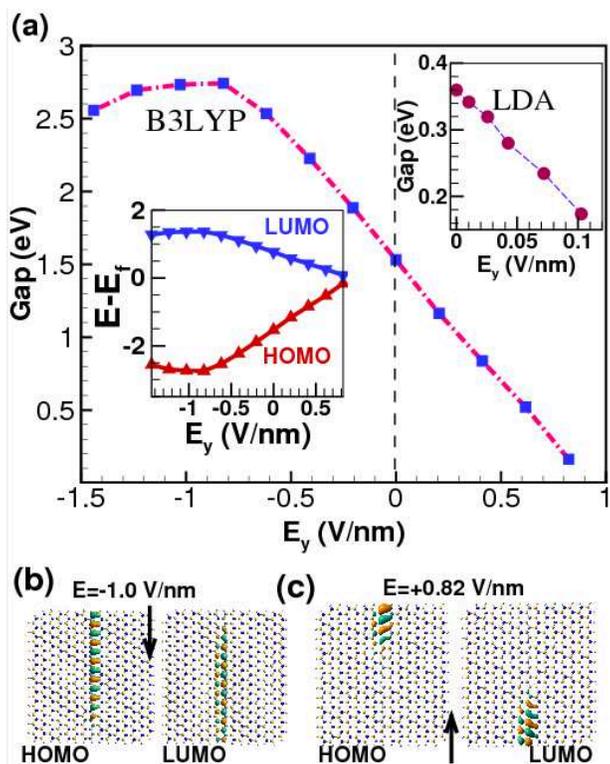}
\caption{(Color online)
 (a) The HOMO-LUMO energy gap for NCB  versus applied electric field using B3LYP.
 Bottom inset  shows the change in HOMO and LUMO versus applied electric field and the top inset
shows the LDA results for  positive electric field. The change in
the HOMO and LUMO orbitals of NCB in the presence of electric field
$E_y=$-1.0 V/nm (b) (+0.82 V/nm (c)). \label{fig:GofE} }
\end{center}
\end{figure*}

 Next we will focus on the NCB system to study its response to an external
electric field. When an electric field in the opposite direction of
the dipole moment  of the NCB is applied,
 the gap becomes smaller, see Fig.~\ref{fig:GofE}(a), and the HOMO (LUMO) is localized much closer to
 the top (bottom), see Fig.~\ref{fig:GofE}(c).
 On the other hand, we found in the previous section that
 the electric polarizability in response to an external field is increased by doping the BNNRs by  carbon atoms.
 This trend can be used to obtain metallic properties in
 doped NCB. Insertion of different numbers of carbon lines as well as
 using  different arrangements  of the atoms near the C-line
 can be a practical tool to tune the electric transmission through the BNNRs and their metallic properties.

An external uniform electric field $E_y$ along the C-line in NCB
causes a potential difference along the C-line. We assume that the
HOMO and the LUMO are effectively centered around $y_{H}$ and
$y_{L}$, respectively. An applied electric field ($E_y$) can
modulate the energy gap as $(y_{L}-y_{H})e\,E_y$, where $e$ is the
elementary charge. The gap can be tuned by the strength and
direction of the electric field. The latter linear behavior is
obtained even using the LDA (see top inset in
Fig.~\ref{fig:GofE}(a)) and is continued (with a slope of
$(y_{L}-y_{H})e$) while the centers of the HOMO and the LUMO are
more or less fixed which is the case even for a rather strong field
of E=+0.82\,V/nm (compare Fig.~3(b) and Fig.~6(c)). The same linear
behavior is found for the reverse electric fields $E_y>-0.8$\,V/nm,
see Fig.~6(a).

On the other hand, if a strong negative field about $-1$\,V/nm is
applied, the dependence is no longer linear. Fig.~6(b) shows that
such a negative field alters the spatial distribution of the
orbitals and elongates both HOMO and LUMO into a non-localized
distribution similar to those found for the BCB or NCN systems.
Therefore around the maximum, $y_{L} \approx y_{H}$ and the energy
gap is independent of the field strength. However for the negative
fields beyond the maximum the energy gap decreases which corresponds
to the localization of the LUMO and the HOMO where $y_{L} < y_{H}$.
This simple picture argues that $(y_{L}-y_{H})e\,E_y$ modulates the
energy-gap.
 Notice that the slope $(y_{L}-y_{H})e$ changes sign passing the maximum
 ($E_y\simeq$ -1\,V/nm), therefore by decreasing the electric field, the gap decreases
when $E_y\succeq$  -1\,V/nm and increases when $E_y \preceq$
-1.0\,V/nm.
 A similar behavior was found for a semi-infinite non-doped BN systems~\cite{21,22}, with
 difference in maximum and energy gap values. The bottom inset in
Fig.~6(a) shows the variation of the HOMO and LUMO sates with the
electric field. We believe that the control of the conductivity by
an external field as we found for the NCB will be of great practical
interest.

 In summary, we studied the electronic properties  of
boron-nitride sheets in the presence of a zigzag  atomic line of
carbon atoms in the middle of the BN sheets.  Electronic
polarization of the BNNRs was found to depend on the local
arrangement of the boron and nitrogen atoms around the carbon line.
Doping with carbon atoms decreases the non-metallic properties of
the BNNRs by reducing the energy gap and increasing the
polarizability. By applying an electric field along the carbon line
we showed that the dipole moment of BNNR induced by the carbon
dopants can be reduced and even eliminated. This also reduces the
energy gap in the electron spectrum  and allows one to control
the conductivity with an external field.\\

\emph{{\textbf{Acknowledgment}}}. We would like to thank J. M.
Pereira and Stefan Goedecker for helpful discussions. This work was
supported by the Flemish Science Foundation (FWO-Vl), the
ESF-EuroGRAPHENE project CONGRAN and the Swiss National Science
Foundation (SNF).

\newpage

\begin{table*}
\caption{Bond lengths in~\AA,~atomic charge in e, HOMO-LUMO band gap
in eV and spontaneous electric dipole moment $P_0$ in D for the four
studied nanoribbons in Fig.~\ref{fig0}. Results for the gap are
given for both B3LYP and LDA (respectively calculated by G09 and
BigDFT).}
\begin{tabular}{ |c| c c c c| c c c c | c c | c |}
 \hline
%  Nanoribbon& $d_{B-N}$  & $d_{B-C}$ & $d_{N-C}$&$d_{C-C}$ & $e_{B}$&  $e_{N}$&  $e_{C-B}$ &$e_{C-N}$& gap:G09 & BigDFT& $P_o$:G09 & BigDFT
  Nanoribbon& $d_{B-N}$  & $d_{B-C}$ & $d_{N-C}$&$d_{C-C}$ & $e_{B}$&  $e_{N}$&  $e_{C-B}$ &$e_{C-N}$& gap:B3LYP & LDA& $P_o$:B3LYP
\\
     \hline
  BN& 1.45 & - & -&-&1.17 &-1.17&  -&-&5.69 & 3.85 &10.8  \\
  NCB     & 1.45 & 1.55& 1.41& 1.40&1.01 & -0.89&  -0.47 &0.32&1.52 & 0.39 &44.24  \\
  BCB     & 1.45 & 1.57 & -&1.46& 1.041&-1.16& -0.28 &-& 1.47 & 0.60 &0.08  \\
  NCN    & 1.45 & -     & 1.43& 1.39&1.15&-0.09 &-&0.13& 1.47 & 0.46 &0.27 \\
  \hline
\end{tabular}
\label{table2}
\end{table*}
\newpage

% The
% simulated system with dimension 3.46$\times$ 3.11 $nm^2$ for a \textbf{pure boron-nitride nanoribbon} (BN).
% Yellow and blue circles indicate boron and nitrogen atoms, respectively, and
% small gray circles at the boundary are hydrogens. Here
% $P_x$=-10.58\,D and $P_y$=-1.98\,D.

%\begin{figure*}
%\begin{center}
%\includegraphics[width=0.8\linewidth]{fig1.eps}
%\caption{(Color online)  (a) Perfect boron-nitride nanoribbon (BN).
%(b) Boron-nitride nanoribbon with a {zigzag C-line in the middle
%surrounded  by  both boron and nitrogen atoms on both sides (BCN).
%(c) Boron-nitride nanoribbons with the C-line surrounded only by
%boron atoms on both sides (BCB). (d) Boron-nitride nanoribbons with
%the C-line surrounded only by nitrogen atoms on both sides (NCN).}
%Vertical red arrows refer to the atomic line of carbon atoms and the
%horizontal black arrows refer to the local \textbf{dipoles for only
%the horizontal bonds.} {The presented figures are only the central
%parts of the simulated systems of size 3.46$\times$ 3.11 $nm^2$
%(Fig.~\ref{fig0}). Numbers refer to the bond lengths in \AA~unit.}
%\label{fig1} }
%\end{center}
%\end{figure*}

\end{document}